\documentclass[3p,twocolomn]{elsarticle}
\usepackage{epsfig}
\usepackage{graphicx}   
\usepackage{dcolumn}    
\usepackage{bm}         
\usepackage{color,multirow,amsmath}
\usepackage{threeparttable}

\newcommand{\etal}{\emph{et al.\@}}\newcommand{\eg}{\emph{e.g.\@} }
\long\def\symbolfootnote[#1]#2{\begingroup%
\def\thefootnote{\fnsymbol{footnote}}\footnote[#1]{#2}\endgroup}

\long\def\symbolfootnotemark[#1]{\begingroup%
\def\thefootnote{\fnsymbol{footnote}}\footnotemark[#1]\endgroup}
\long\def\symbolfootnotetext[#1]#2{\begingroup%
\def\thefootnote{\fnsymbol{footnote}}\footnotetext[#1]{#2}\endgroup}

\begin{document}
\title{Neutron-induced \textcolor{black}{dpa,} transmutations, gas production, and helium embrittlement of fusion materials}

\author{M.R. Gilbert\symbolfootnotemark[1]}
\author{S.L. Dudarev}
\author{D. Nguyen-Manh}
\author{S. Zheng}
\author{L.W. Packer}
\author{J.-Ch. Sublet}
\address{EURATOM/CCFE Fusion Association, Culham Centre for Fusion Energy,
Abingdon, Oxfordshire OX14 3DB, UK.}
\begin{abstract}
In a fusion reactor materials will be subjected to significant fluxes of high-energy neutrons. As well as causing radiation damage, the neutrons also initiate nuclear reactions leading to changes in the chemical composition of materials (transmutation). Many of these reactions produce gases, particularly helium, which cause additional swelling and embrittlement of materials. This paper investigates, using a combination of neutron-transport and inventory calculations, \textcolor{black}{the variation in displacements per atom (dpa) and helium production levels as a function of position within the high flux regions of a recent conceptual model for the `next-step' fusion device DEMO. Subsequently, the gas production rates are used to provide revised estimates, based on new density-functional-theory results, for the critical component lifetimes associated with the helium-induced grain-boundary embrittlement of materials. The revised estimates give more optimistic projections for the lifetimes of materials in a fusion power plant compared to a previous study, while at the same time indicating that helium embrittlement remains one of the most significant factors controlling the structural integrity of fusion power plant components.}
\end{abstract}

\date{\today}
\maketitle

\symbolfootnotetext[1]{Corresponding author email: mark.gilbert@ccfe.ac.uk}

\section{Introduction}

In both the currently planned ITER device and in the next-step fusion devices, one of the key outstanding issues lies in the understanding of how materials are affected by the high fluxes of neutrons produced by the fusion plasmas. Not only do the incident neutrons cause atomic displacements within materials surrounding the plasma, leading to defect accumulation, but they can also initiate non-elastic nuclear reactions that cause the atoms of a material to be altered (transmuted), leading to a change in the structure and behavior of components. Even more problematic are the subset of the possible nuclear reactions that produce gas particles (helium and hydrogen). Helium (He) in particular, with its low reactivity, can persist in materials over long periods of time, leading to accumulation in existing cracks or at grain boundaries, which can result in swelling or embrittlement.

\begin{figure}[t]
\begin{centering}
{\includegraphics[angle=0,width=0.45\textwidth,clip=true,trim=0cm 0cm 0cm
0cm]{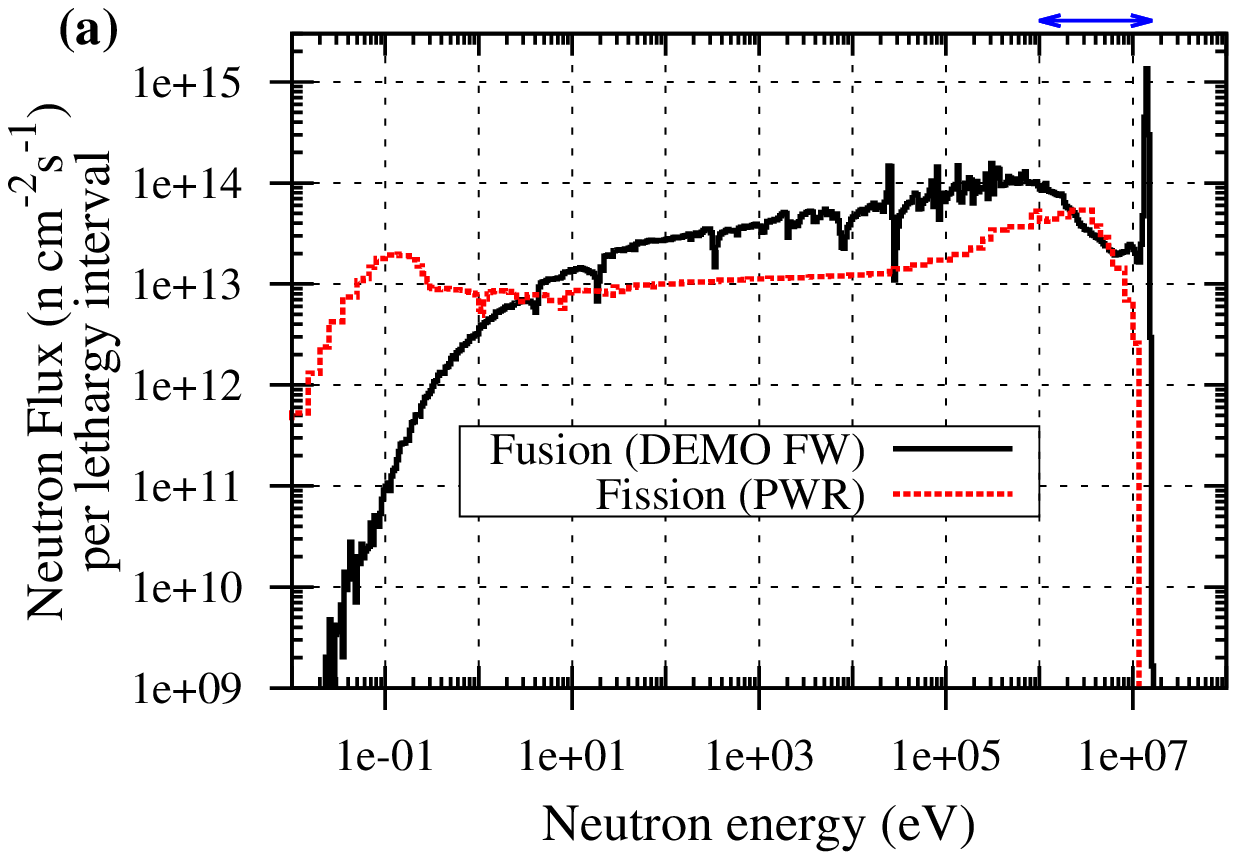}}
{\includegraphics[angle=0,width=0.45\textwidth,clip=true,trim=0cm 0cm 0cm
0cm]{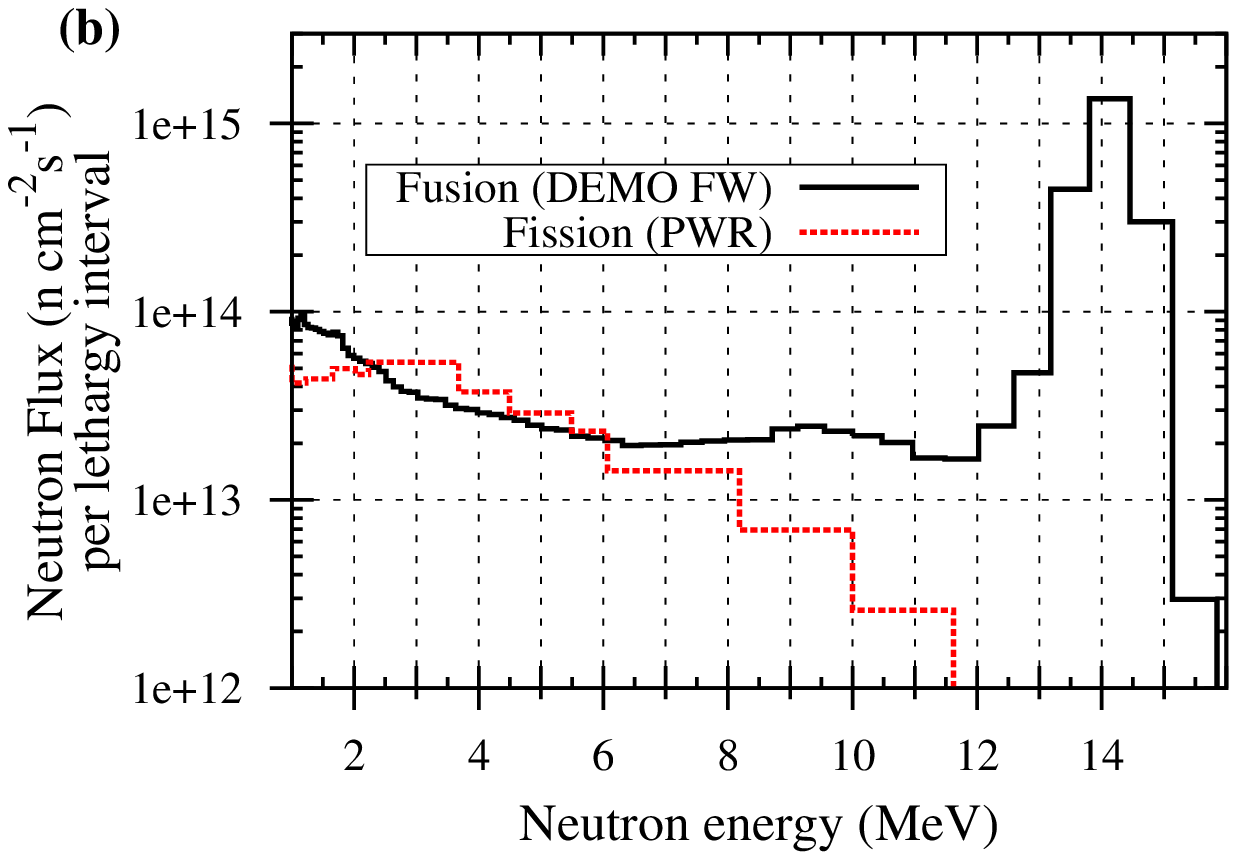}}
\caption{\label{fission_fusion_spectra}\textcolor{black}{Comparison of the neutron-energy spectra in fission and fusion reactors. Shown is the average neutron spectrum in the fuel-assembly of a PWR reactor and the outboard equatorial FW spectrum for the DEMO model in figure~\ref{mcnp_model_demo}. In (a) the full energy range of the spectra are shown on an eV scale -- down to the thermal energies, while in (b) only the portion of the spectra above 1~MeV are shown on a MeV energy-scale, which covers the range indicated by the arrow in (a) (note also the change from logarithmic to linear scale). The spectra are measured in energy bins of varying width and so are plotted as step curves. The fusion spectrum has significant fluxes of neutrons above 12~MeV, leading to both an increase in probability for threshold nuclear reactions and higher dpa levels. Notice also the wide peak in the fusion spectrum around \(E=14\)~MeV, which is caused by Doppler broadening of the neutron spectra associated with the thermal motion of plasma ions with a \(\sqrt{k_BT\cdot E}\) dependence, where \(T\) is plasma temperature, and results in a significant number of neutrons with energies of 15~MeV or higher.}}
\end{centering}
\end{figure}

Since many of these gas-producing nuclear reactions have threshold energies, gas production and any subsequent swelling is more of a concern in fusion compared to fission because of the larger fraction of neutrons at higher energies and higher overall flux (see for example, figure~\ref{fission_fusion_spectra}, which compares the neutron flux per lethargy interval\footnote{lethargy interval is a commonly used measure for
spectra of this type, and is equal to the natural logarithm of the ratio of a given energy-interval's
upper bound to its lower bound.} as a function of energy for the 3.0~GWt [gigawatts  of thermal power] DEMO concept to a 3.8~GWt fission reactor).

\textcolor{black}{This paper describes the latest results from integrated studies for a conceptual design of DEMO---the demonstration fusion power-plant---combining neutron-transport simulations to define the variation in irradiation environment, inventory calculations to predict the transmutation of materials and build-up of impurities, and simple, atomic-level modeling of the consequences associated with, in this study, the production of helium-gas as an impurity.
In particular, we focus on recent advances in calculation of displacements per atom (dpa) rates for neutron-irradiated materials, which are used as a measure of irradiation dose, and also revisit our earlier calculations~\cite{gilbertetal2012} of the estimated lifetimes associated with the helium-embrittlement of grain-boundaries in light of new understanding from density-function-theory simulations.}

\section{Neutron transport and inventory simulations}

In an earlier work~\cite{gilbertsublet2011} we focused on the transmutation response of various materials under identical first wall (FW) conditions. In the present studies we go further and investigate the response of materials as a function of position within a fusion device, with particular emphasis on how helium production rates change.

\subsection{Geometry dependence of neutron flux and energy spectrum}

\begin{figure*}[p]
\begin{centering}
{\includegraphics[width=0.8\textwidth,clip=true,trim=0cm 0cm 1.1cm
0cm]{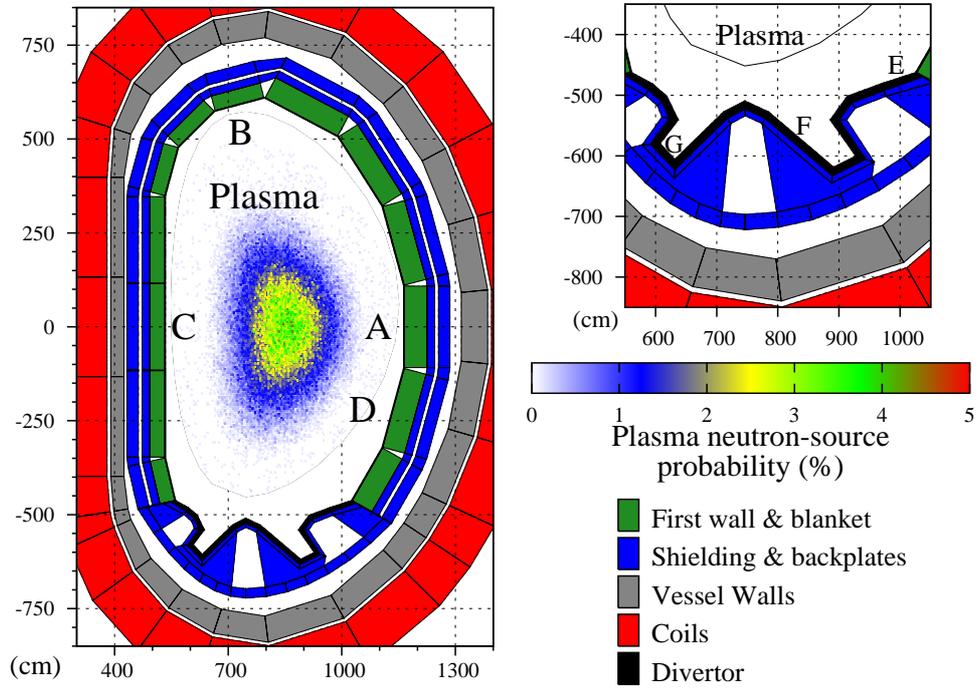}}
\caption{\label{mcnp_model_demo}The simplified, homogeneous DEMO model used in MCNP simulations to obtain neutron fluxes and spectra. \textcolor{black}{The full model shown on the left also includes a sampled location probability distribution for the neutrons generated in the plasma.}}
\end{centering}
\end{figure*}

\begin{figure}[p]
\begin{center}
{\includegraphics[angle=0,width=0.4\textwidth,clip=true,trim=-3cm 2cm 0cm
2cm]{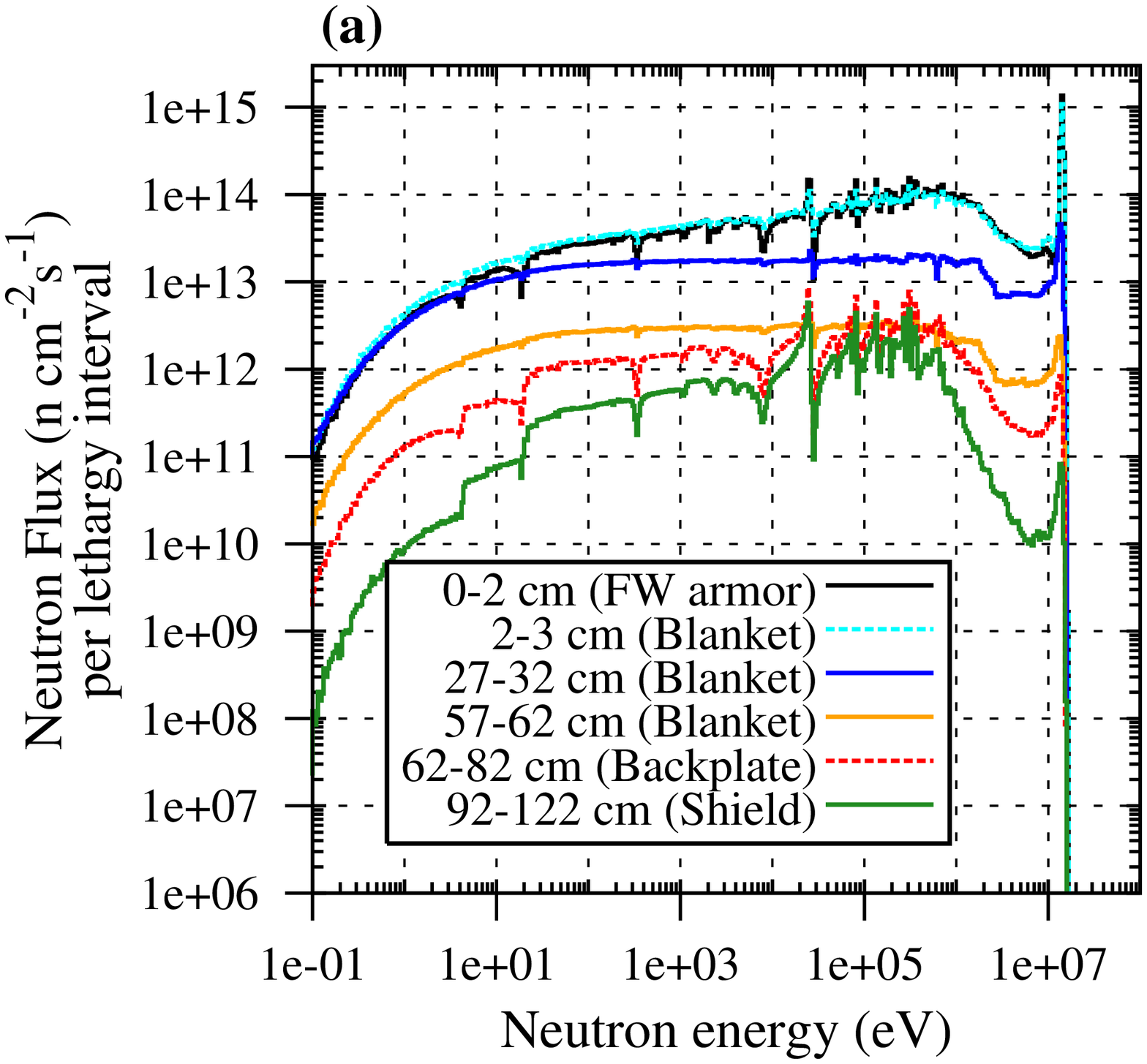}}
{\includegraphics[angle=0,width=0.4\textwidth,clip=true,trim=-3cm 2cm 0cm
2cm]{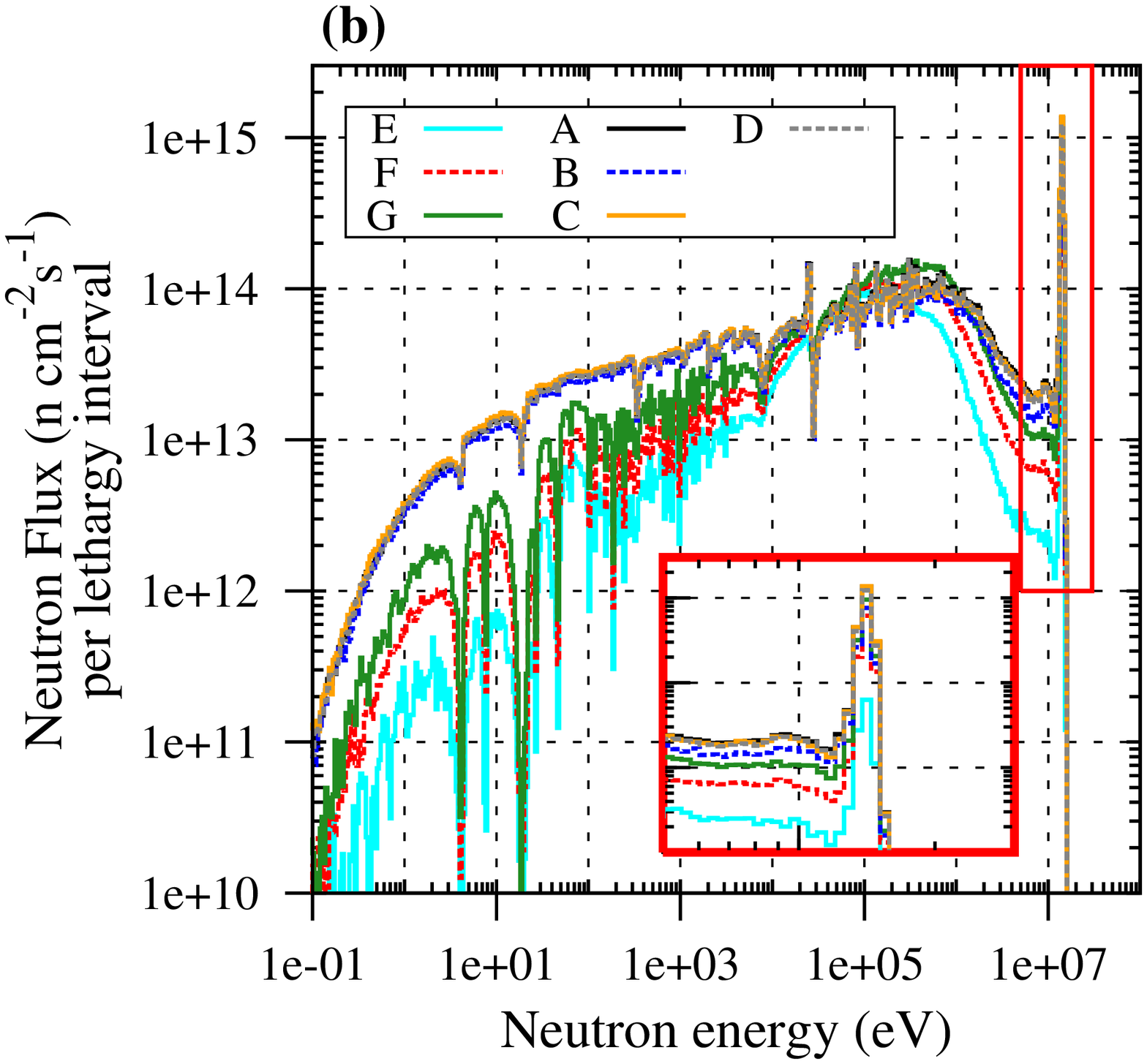}}
\caption{\label{demo_spectra}Comparison of the neutron-energy spectra in DEMO; (a) as a function of distance from the plasma-facing surface of the containment vessel at the equatorial position (A) in figure~\ref{mcnp_model_demo}; and (b) \textcolor{black}{in the 2~cm armor layer as a function of position (A--D in the FW and E--G in the divertor -- see figure~\ref{mcnp_model_demo}).}}
\end{center}
\end{figure}

Neutron spectra have been calculated for different regions of a recent DEMO design, proposed by CCFE in 2009~\cite{zheng2011} (see figure~\ref{mcnp_model_demo}), using the neutron-transport code MCNP~\cite{forrestmcnp}. The model is highly simplified, with only the major structures included, and with homogeneous material compositions. Several of the spectra calculated for this model are shown in figure~\ref{demo_spectra}.

Figure~\ref{demo_spectra}b shows that both the energy profile and the fluxes of the neutron spectra vary dramatically as a function of distance from the plasma-facing surface into the equatorial region of the FW at A in figure~\ref{mcnp_model_demo}. Not only does the neutron spectrum become heavily moderated as the distance from the plasma increases, but the total flux also falls greatly -- from \(8.25\times 10^{14}\)~n~cm\(^{-2}\)~s\(^{-1}\) in the 2~cm FW armor at position A to \(3.9\times 10^{13}\)~n~cm\(^{-2}\)~s\(^{-1}\) in the final 5~cm of the blanket.

Within the divertor  the neutron flux and spectrum also show significant variation as a function of position (see figure~\ref{demo_spectra}b). At point E in figure~\ref{mcnp_model_demo}, the total flux in the 2~cm layer of pure W divertor armor is approximately twice as high as that  at G (\(7.1\times 10^{14}\)~n~cm\(^{-2}\)~s\(^{-1}\) at E compared to \(3.6\times 10^{14}\)~n~cm\(^{-2}\)~s\(^{-1}\)  at G).

\subsection{Spectral influence on transmutation and helium production}

The calculated neutron spectra and total fluxes  for the DEMO model have been used in the inventory code FISPACT~\cite{forrestfispact2007} to simulate the burn-up (transmutation) of various materials relevant to design. FISPACT requires an external library of reaction cross sections and decay data, and here we have employed the 2003 version of the European Activation File (EAF)~\cite{forresteasy2003}. Note that, for W, we employ the self-shielding correction-factors obtained in~\cite{gilbertsublet2011} to the appropriate reactions used within FISPACT. Below we discuss the helium production rates in various fusion-relevant materials.

The gas production from iron (Fe), as the primary constituent of steels, will be a major factor in determining the lifetime of near-plasma components in fusion reactors. Chromium (Cr), which forms around 10\% of the composition of the reduced activation steels being proposed for fusion application, has a very similar transmutation profile to Fe~\cite{gilbertsublet2011}.

\begin{figure*}[t]
\begin{centering}
{\includegraphics[height=0.4\textwidth,clip=true,trim=0cm 1cm 0cm
1cm]{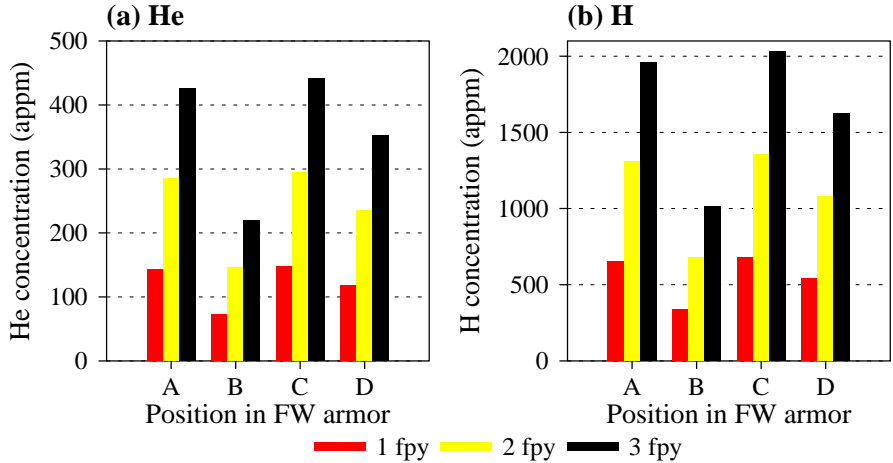}}
\caption{\label{fe_gas_production_FW}Variation in the (a) He, and (b) H, concentrations in pure Fe as a function of time (in full power years (fpy)) for the neutron spectra in different first wall (FW) armor positions in DEMO -- see figure~\ref{mcnp_model_demo}.}
\end{centering}
\end{figure*}

\begin{figure}[t]
\begin{minipage}[b]{0.49\linewidth}
\begin{center}
{\includegraphics[width=1.0\textwidth,clip=true,trim=0.0cm 0cm 0.0cm
0.0cm]{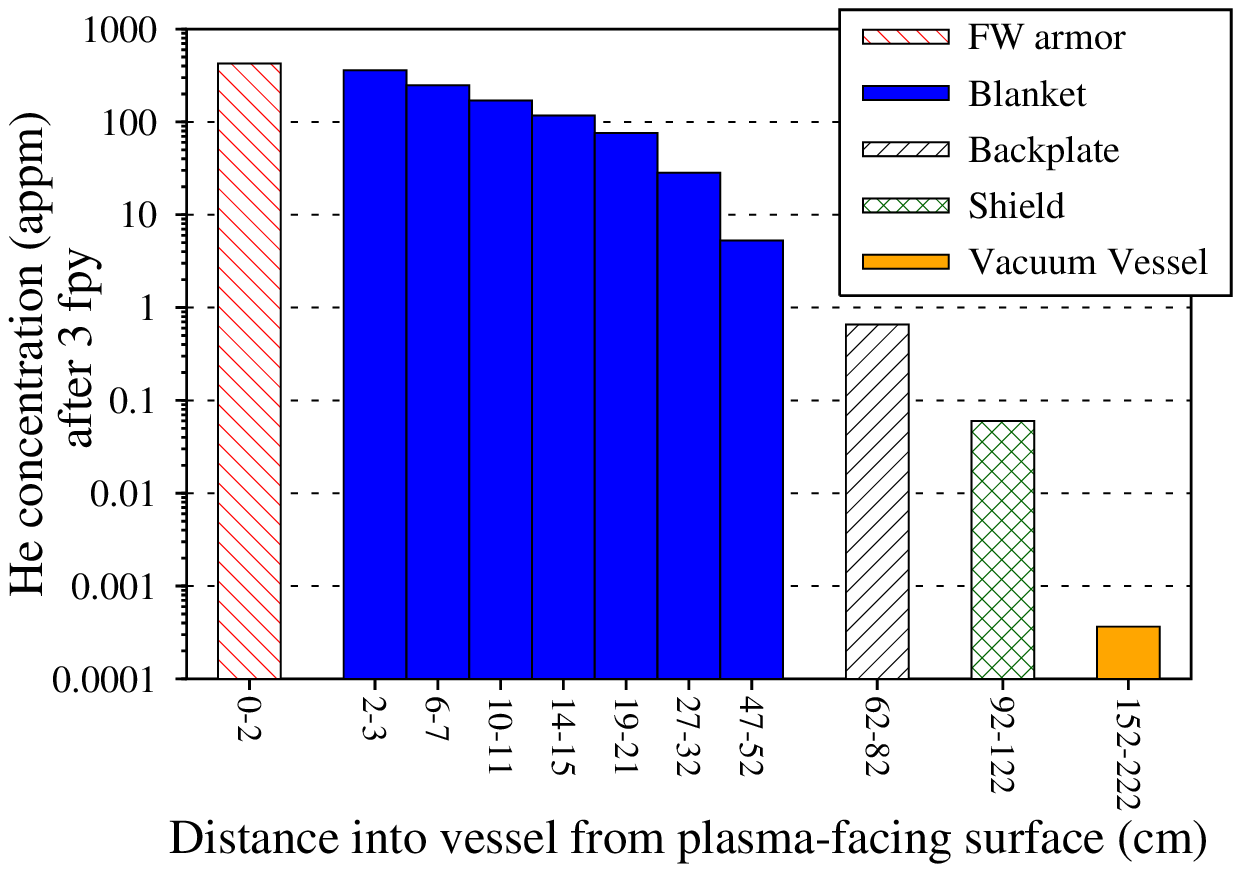}}
\caption{Variation in He concentration in pure Fe after a 3~fpy irradiation as a function of distance from the plasma-facing surface of DEMO at (A) in figure~\ref{mcnp_model_demo}. Note the logarithmic scale of the concentration axes.}
\label{fe_gas_production_depth}
\end{center}
\end{minipage}
\begin{minipage}[b]{0.02\linewidth}
\quad
\end{minipage}
\begin{minipage}[b]{0.45\linewidth}
\begin{center}
{\includegraphics[angle=0,width=1.0\textwidth,clip=true,trim=0cm -0.1cm 2cm
-0.5cm]{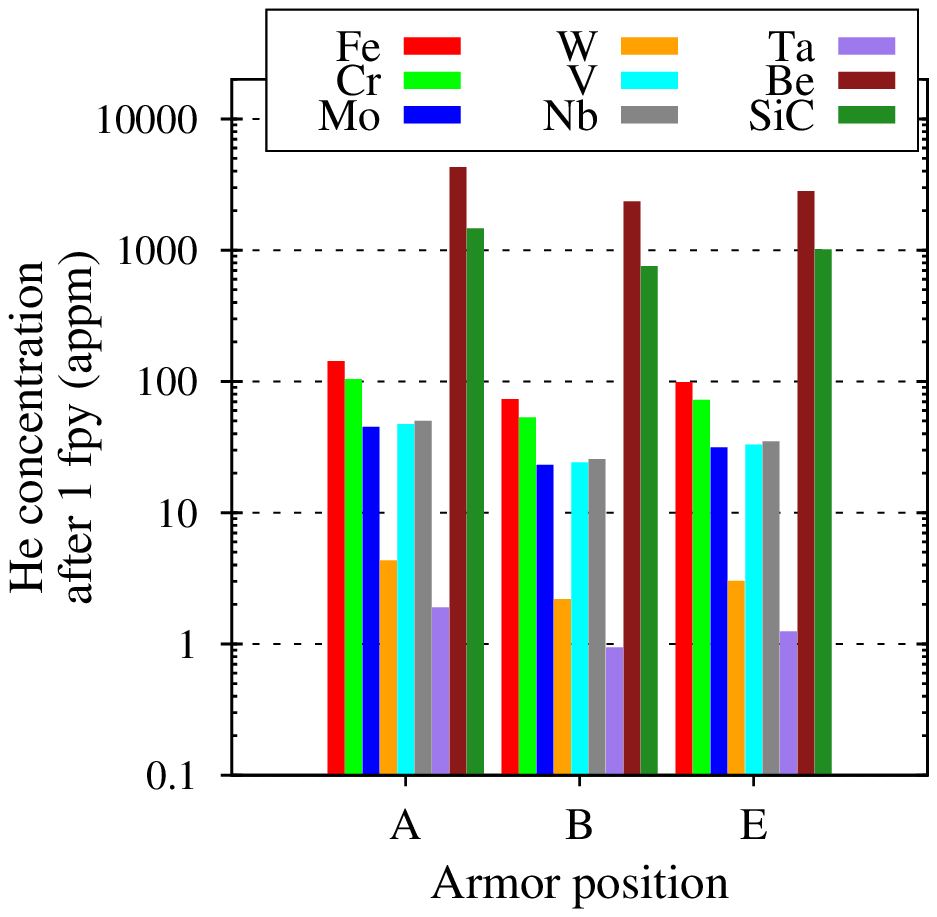}}
\caption{Variation in He concentration in various pure materials after a 1~fpy irradiation in different armor spectra. The positions refer to those indicated in figure~\ref{mcnp_model_demo}.}
\label{multi_he_production_armour}
\end{center}
\end{minipage}
\end{figure}

Figure~\ref{fe_gas_production_FW}a shows how the concentration of He produced under irradiation varies as a function of time and of position on the FW armor of the DEMO model. For comparison, figure~\ref{fe_gas_production_FW}b shows the equivalent hydrogen (H) production levels from Fe, which are approximately 5 times higher. The irradiation times shown here and elsewhere in this work are DEMO full-power years (fpy). In Cr, the helium production rates are 20-25\% lower than in Fe (see for example figure~\ref{multi_he_production_armour}), but the variations as a function of time and position are similar.

At position B in the FW, the significant drop in gas production from Fe shown in figure~\ref{fe_gas_production_FW} is due to a combination of a reduced total flux (\(\sim7\times 10^{14}\)~n~cm\(^{-2}\)~s\(^{-1}\) vs. \(\sim8\times 10^{14}\)~n~cm\(^{-2}\)~s\(^{-1}\) at the other positions) and a lower proportion of neutrons above the threshold energies for the gas producing reactions (\eg, the (\(n,\alpha\)) reaction on \(^{56}\)Fe, which is the main production route for He, has a threshold of approximately 3.7~MeV~\cite{forresteasy2003}) -- see figure~\ref{demo_spectra}b.

As a function of distance from the plasma-facing surface, inventory calculations reveal that the changes in the neutron irradiation conditions (figure~\ref{demo_spectra}) cause the He production levels to fall significantly in Fe. For example, after 3 years under DEMO full-power conditions, the He concentration is around 400 atomic parts per million (appm) in the FW armor, but has barely reached 1~appm in the 20~cm backplate behind the blanket (see figure~\ref{fe_gas_production_depth}).

Note that the neutron spectra calculated by MCNP do not take into account time-dependent compositional changes in materials, such as those that will take place in the blanket as the Li is depleted during tritium breeding. It is possible to investigate these processes by using an inventory code, such as FISPACT, to periodically update the material compositions in an MCNP calculation. Packer~\etal~\cite{packeretal2011} have recently applied such a methodology to investigate how the tritium-breeding inventory evolves in the DEMO blanket.

Tungsten (W) will be present throughout a typical reactor vessel as small concentrations in most steels and will be used in an almost pure form in the high heat-flux divertor regions due to its high melting-point, thermal-conductivity, and resistance to sputtering and erosion~\cite{nemotoetal2000}. In many reactor designs W is also considered for the FW armor layer~\cite{maisonnieretal2005}.

\begin{figure*}[t]
\begin{centering}
{\includegraphics[height=0.4\textwidth,clip=true,trim=0cm 0.4cm 0cm
1cm]{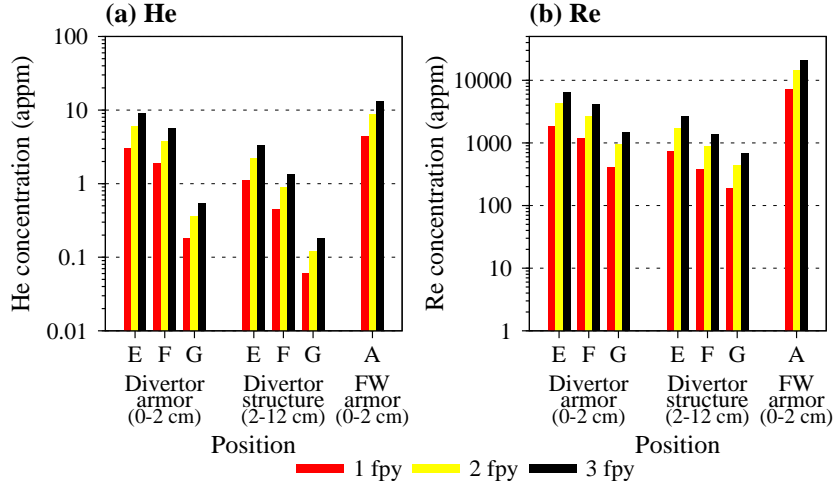}}
\caption{\label{W_transmutation_results}Variation in concentrations of (a) He, and (b) Re, produced in pure W under neutron irradiation as a function of position (and distance from plasma-facing surface) in different regions of the DEMO design (labels as indicated in figure~\ref{mcnp_model_demo}).}
\end{centering}
\end{figure*}

For the present DEMO model the FW armor layer is very thin, meaning that it is almost transparent to neutrons, and so it is realistic to assume that the irradiation conditions found for FW on DEMO  in the previous section are very close to those that would be obtained if the FW were W instead. The results from inventory calculations (figure~\ref{W_transmutation_results}) reveal that the concentration of He produced from pure W is much lower than from Fe, but can also vary significantly, even within the same layer of the divertor.

The amount of He produced in the divertor armor (shown in figure~\ref{mcnp_model_demo}) after the first 3 years of irradiation varies from 9~appm at position E to less than 1~(one)~appm at G -- an order of magnitude difference (figure~\ref{W_transmutation_results}a). In the divertor structure behind the armor the variation is similar, albeit at a systematically lower level. For H the variations with position are similar but the absolute concentrations are roughly twice those found for He. The higher flux in the FW armor causes the gas concentration to be somewhat greater than those observed for the divertor (see figure~\ref{W_transmutation_results}a).

Perhaps of greater significance for W are the variations in rhenium (Re) concentration between the FW armor and the divertor (figure~\ref{W_transmutation_results}b), because of the potential for the formation of \(\sigma\)-phase precipitates (with Os)~\cite{cottrell2004}. After 3 years in the FW armor at A, Re reaches a concentration of around 20000~appm (2~atomic~\%), which is broadly in line with the previous findings in~\cite{gilbertsublet2011}. However, in the divertor armor at position E, Re only reaches a concentration of around 6400~appm (less than 1~atomic~\%) on the same timescale.

\textcolor{black}{Figure~\ref{multi_he_production_armour} compares the He production from several different materials, including Fe and W, under different armor conditions after 1~fpy. It is immediately obvious that some materials, such as Be and SiC, have significantly higher production rates than the benchmark of Fe. For example, in the outboard equatorial FW armor (A in figure~\ref{mcnp_model_demo}) Be, which is a key neutron moderator in tritium breeding blanket designs, produces around 4300~appm in 1~fpy, compared to the \(\sim140\) from Fe and \(\sim4\) from W. Other transition metals, such as Cr, Mo, V, and Nb have similar production levels to Fe (although Fe is the highest), while Ta has production rates as low as W.}


\subsection{Variation in dpa}

\textcolor{black}{A common method of interpretation for neutron spectra and fluxes is to convert a given set of irradiation conditions into an integrated quantity known as displacements per atom (dpa). This is particularly done to inform materials modelers and experimentalists about the irradiation dose at the atomic level. In our previous publication~\cite{gilbertetal2012}, we obtained dpa rates for different materials under various conditions predicted for the DEMO model using the modified Kinchen-Pease method of Norgett, Robinson, and Torrens (NRT)~\cite{norgettetal1975}. Specifically, the energy-dependent total dpa cross sections were obtained from data contained in the European Fusion File (EFF) 1.1 and processed using the NJOY~\cite{njoy1994} nuclear data processing system. Below we highlight two particular aspects of dpa evaluations, which should be carefully considered when using such data.}

\begin{figure}[t]
\begin{minipage}[b]{0.49\linewidth}
\begin{center}
{\includegraphics[width=1.0\textwidth,clip=true,trim=1cm 0cm 1.5cm
-0.2cm]{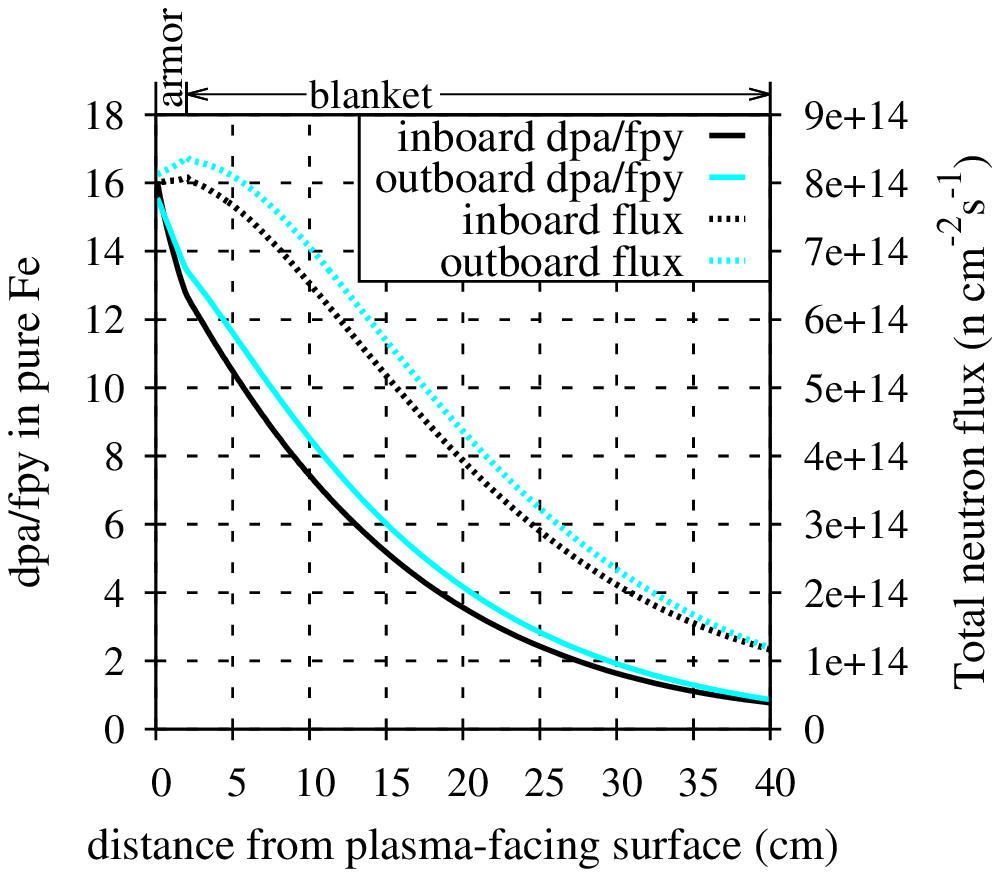}}
\caption{\textcolor{black}{Variation of both the dpa/fpy in Fe and total neutron flux with distance from the plasma-facing surface into the outboard and inboard equatorial regions at (A) and (C), respectively, in fig.~\ref{mcnp_model_demo}.}}
\label{dpa_with_depth}
\end{center}
\end{minipage}
\begin{minipage}[b]{0.02\linewidth}
\quad
\end{minipage}
\begin{minipage}[b]{0.45\linewidth}
\begin{center}
{\includegraphics[angle=0,width=1.0\textwidth,clip=true,trim=1.7cm 0cm 1.4cm
0cm]{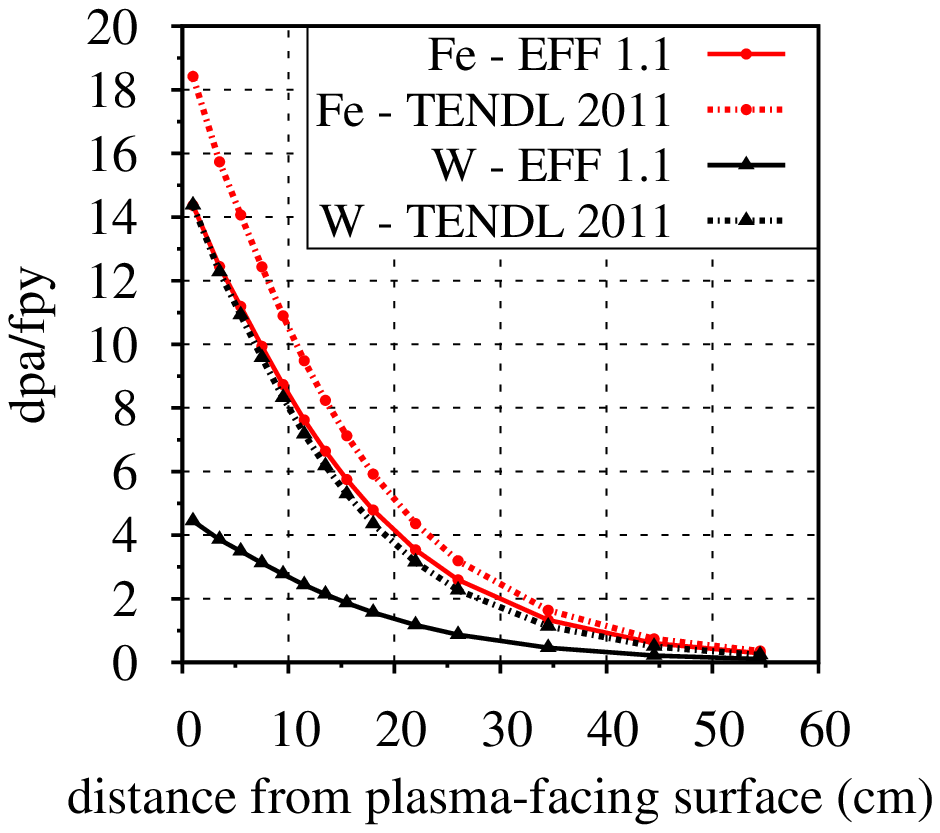}}
\caption{\textcolor{black}{Comparison of dpa/fpy values for pure Fe and W using two alternative nuclear data libraries. Results are given as a function of distance from the surface into the outboard equatorial FW (A in figure~\ref{mcnp_model_demo}).}}
\label{dpa_IvsII}
\end{center}
\end{minipage}
\end{figure}

\textcolor{black}{Firstly, in figure~\ref{dpa_with_depth} we compare the variation in dpa/fpy in pure Fe to the equivalent total neutron flux as a function of distance from the plasma-facing surface into the equatorial regions of the inboard and outboard FW of the DEMO model (A and C in figure~\ref{mcnp_model_demo}, respectively). In both cases, the inboard fluxes and dpa/fpy values are generally lower than the outboard values at the same distance from the plasma-facing surface. However, it is also clear that while dpa/fpy is always decreasing with increasing distance, the same is not true of the total flux, which actually increases initially with distance from the plasma due to neutron multiplication in the first few centimeters of the armor and blanket. Thus, using dpa as a measure of damage in this case would hide the fact that certain regions of the FW are experiencing a higher total flux of neutrons.}

\textcolor{black}{Recently the inventory code FISPACT has been extensively updated (now called FISPACT-II~\cite{subletetal2012}) and, specifically, can calculate dpa rates directly using the same NRT method as before, but with the latest nuclear data libraries. Figure~\ref{dpa_IvsII} compares our original calculations from NJOY using EFF~1.1 (used in figure~\ref{dpa_with_depth} and in~\cite{gilbertetal2012}) for pure Fe and W to new results obtained from FISPACT-II using the TENDL-2011 library. The graph shows the variation in dpa/fpy as a function of distance from the surface into the outboard equatorial FW at A in figure~\ref{mcnp_model_demo}. It is immediately obvious that, for W in particular, there has been a dramatic change in the results. Whereas with the EFF~1.1, the dpa/fpy in pure W in the FW armor was around 4.5 dpa/fpy, with the new evaluation the values are more like 14.5 dpa/fpy -- a factor of 3 increase. The change in results for Fe are less extreme, but still non-negligible. These findings highlight the need for caution when using dpa as a measure of irradiation exposure, particularly in cases where the nuclear data maybe ill-defined, for example due to a lack of supporting experimental data, and therefore subject to variation.}

\section{Modeling of He-induced grain-boundary embrittlement}
The calculations described in the foregoing section produce quantitative estimates of the He production rates under any neutron-irradiation conditions. Using a simple model for He segregation to boundaries, we can use these results to estimate the timescales required to produce a sufficient amount of He to embrittle materials. \textcolor{black}{Full details of this model are given in~\cite{gilbertetal2012}. The model uses density-functional-theory (DFT) to calculate the energy associated with inserting He into the system, and experimentally measured surface energies to estimate the critical concentrations of He that can be accommodated at grain boundaries (GBs) before failure. Specifically, the model neglects traps and obstacles in the interior of grains, and assumes that all of the helium generated in the bulk by transmutation can migrate to the boundaries. In this sense the model is most valid in the limit of small grain size, where obstacles will have a smaller effect on the trajectory of He atoms to the GBs. We relate the critical concentration \(\nu_{\text{He}}^c\) in the grain boundary plane to the critical bulk concentration \(G_{\text{He}}^c\) via}
\begin{equation}\label{concentration}
\textcolor{black}{G_{\text{He}}^c=3\nu_{\text{He}}^c/an,}
\end{equation}
where \(a\) is the characteristic grain size (we can assume either cubic or spherical grains without loss of generality), and \(n\) is the atomic density of the material in cm\(^{-3}\). Table~\ref{time_to_critical} shows \(\nu_{\text{He}}^c\) and \(G_{\text{He}}^c\) values for the body-centred-cubic (bcc) transition metals, Fe, Cr, Mo, W, V, Nb, and Ta, for a small assumed grain size of 0.5~\(\mu\)m. From equation~\ref{concentration} we observe that the critical bulk concentrations decrease linearly with increasing grain size, and therefore the predictions obtained from the model are very sensitive to the choice of \(a\) -- although, as stated above, the model is most valid in the limit of small grain size.

In~\cite{gilbertetal2012}, the critical GB concentrations \(\nu_{\text{He}}^c\) were estimated by equating the surface energies \(\varepsilon_{\text{surf}}\) to the energy of solution for a helium atom \(E_{\text{He}}^{\text{sol}}\) in a given material via:
\begin{equation}
\textcolor{black}{E_{\text{He}}^{\text{sol}}\nu^c_{\text{He}}\approx2\varepsilon_{\text{surf}},}
\end{equation}
with \(\varepsilon_{\text{surf}}\) data taken from a database of experimental values reported in~\cite{vitosetal1998}, and average \(E_{\text{He}}^{\text{sol}}\) values obtained from various publications, but specifically from~\cite{willaimefu2006} in the case of the bcc transition metals.
However, strictly speaking, \(E_{\text{He}}^{\text{sol}}\) is associated with a two-step process: that of firstly creating a vacancy-like atomic configuration at the boundary, and then inserting the He atom into it. In a more realistic scenario, the vacancy-like site will already exist -- either as a natural consequence of the mismatch in the orientation of neighbouring grains, or as a result of earlier irradiation damage -- and so this vacancy creation energy should not be included. In effect, the GB can accommodate more He (in pre-existing spaces) than it would otherwise do in the limit of a perfectly aligned boundary with no damage (very rare).

Using this new understanding, we have performed DFT calculations to obtain the energy associated with inserting a He atom into a vacancy site \(E^{\text{insrt}}_{\text{He}}\). These calculations were performed using the Vienna ab-initio simulation package (VASP) \cite{kresse96a,kresse96b} within  the generalized gradient approximation (GGA) with the Perdew-Burke-Emzerhof exchange and correlation functional \cite{pbe1996}. Solution of the Kohn-Sham equations has been carried out self-consistently using a plane-wave basis set with an energy cutoff of 400~eV and with the projector augmented wave (PAW) pseudo-potentials. It is important to emphasise that for all the bcc transition metals, the inclusion of semi-core electrons through the use of pseudo-potentials is important for predicting accurately the defect formation energies~\cite{willaimeetal2000,dnm2006}. Table~\ref{time_to_critical} shows the calculated values of \(E^{\text{insrt}}_{\text{He}}\), and the equivalent \(E_{\text{He}}^{\text{sol}}\) used in~\cite{gilbertetal2012}, for the bcc metals considered.

Table~\ref{time_to_critical} compares the critical boundary \(\nu_{\text{He}}^c\) and bulk \(G_{\text{He}}^c\) He concentrations using the two alternative energies, as well as the approximate time taken \(t^c_{\text{He}}\) to reach each of the \(G_{\text{He}}^c\) values under the FW armor conditions in the outboard equatorial region of the DEMO model. 
As expected, the \(\nu_{\text{He}}^c\) estimates  are higher in the present study, leading to correspondingly higher \(G_{\text{He}}^c\) and \(t^c_{\text{He}}\) values. Note that in both studies, the \(\nu_{\text{He}}^c\), appear to be in reasonable agreement with experimental findings -- particularly for W. Gerasimenko~\etal~\cite{Gerasimenkoetal1998} found that GBs in helium-irradiated W came apart at He concentrations of the order of \(10^{14}\)--\(10^{15}\)~cm\(^{-2}\).

More important than the absolute values, which will be very sensitive to the assumptions made, particularly the grain size and omission of the effect of migration barriers, are the trends in different materials. For example, in~\cite{gilbertetal2012} it was observed that Be, with its high He production rates, had the shortest critical lifetimes (new DFT calculations for Be have not been performed here). In table~\ref{time_to_critical} the highest critical bulk concentrations and lifetimes are associated with W and Ta, primarily because He is produced at such a low level in these materials (although they also have the highest surface energies -- see~\cite{gilbertetal2012}). On the other hand, the results for Fe suggest that, even with the revised estimates, the issue of He-embrittlement of GBs could become an issue on timescales similar to the required lifetimes (5--10 years) of components in fusion reactors.

{\begin{table*}[t]
\center\begin{threeparttable}\caption{\label{time_to_critical}\textcolor{black}{Table of calculated critical boundary densities \(\nu^c_{\text{He}}\), critical bulk concentrations \(G^c_{\text{He}}\), and approximate critical lifetimes \(t^c_{\text{He}}\) (in DEMO first-wall full-power time) for He in various elements. Estimates based on the energy of solution method in~\cite{gilbertetal2012} are compared to the present work using insertion energies for He. Assumed grain size of 0.5~\(\mu\)m.}}

\begin{tabular}{c|cccc|cccc}\hline
\multirow{4}{*}{Element}& \multicolumn{4}{c}{Original work in~\cite{gilbertetal2012}} \vline& \multicolumn{4}{c}{Present study} \\\cline{2-9}\vspace{-2.5ex}\\
&  \(E^{\text{sol}}_{\text{He}}\)&\(\nu^c_{\text{He}}\)&\(G^c_{\text{He}}\)& \(t^c_{\text{He}}\)(FW) &\(E^{\text{insrt}}_{\text{He}}\)&\(\nu^c_{\text{He}}\)&\(G^c_{\text{He}}\)& \(t^c_{\text{He}}\)(FW)\\
&(eV)&(cm\(^{-2}\))&(appm)&(fpy)&(eV)&(cm\(^{-2}\))&(appm)&(fpy)\\
\hline\hline
Fe &4.34&\(6.90\times 10^{14}\)&488.0&4 &2.77&\(1.08\times 10^{15}\)&764.6&6\\
Cr&5.2&\(5.52\times 10^{14}\)&397.8&4&2.68&\(1.07\times 10^{15}\)&771.9&8\\
Mo&4.65&\(8.05\times 10^{14}\)&753.2&18&1.91&\(1.96\times 10^{15}\)&1833.8&46\\
W&4.77&\(9.16\times 10^{14}\)&871.5& 326&1.61&\(2.71\times 10^{15}\)&2582.1&700+\\
V&4.81&\(6.75\times 10^{14}\)&560.5&12&2.3&\(1.41\times 10^{15}\)&1172.2&37\\
Nb&4.55&\(7.41\times 10^{14}\)&800.1&17&1.6&\(2.11\times 10^{15}\)&2275.2&49\\
Ta&4.82&\(7.77\times 10^{14}\)&841.3&216&1.69&\(2.22\times 10^{15}\)&2399.4&700+\\
\hline\hline

\end{tabular}

\end{threeparttable}
\end{table*}

\section{Summary}
\textcolor{black}{This paper described the latest results from an integrated computational study for the neutron-irradiation conditions in a fusion-reactor device. The main points in this paper are:}

\begin{itemize}
\item\textcolor{black}{ The MCNP calculations conducted with the conceptual CCFE design of a DEMO fusion reactor show that the neutron irradiation conditions can vary significantly as a function of position within the reactor. Even in the same component, the flux can change dramatically over short distances. As a function of distance from the plasma-facing surface of the FW, the flux drops by several orders of magnitude and the energy spectrum becomes considerably softer.}
\item \textcolor{black}{The FISPACT-inventory calculations reveal how the variation in conditions alters the rates of production of impurities from transmutation reactions. In particular, helium concentrations fall by many orders of magnitude from the thin FW-armor layer to the outer regions of the vessel, such as the shield. In Fe, for example, the production of He of up to 140~appm per fpy in the FW-armor is likely to be significant because concentrations in the range of 400~appm are known to cause a change in the fracture behavior of neutron-irradiated steels compared to those exposed to neutrons alone~\cite{yamamotoetal2006}. However, such considerations will quickly become unimportant in regions further from the plasma, since even within the blanket the He production rate in Fe falls below 10~appm per fpy. Furthermore, in some materials, such as W, the He production levels in the bulk (as opposed to direct implantation at the plasma-facing surface) are probably too low to have any impact on component lifetime.}
\item\textcolor{black}{ An integrated quantity, such as dpa, can sometimes obscure the true variation in irradiation environment, and the sensitivity of these dpa calculations to nuclear data, as highlighted in this work, casts doubt on the suitability of dpa as the ubiquitous measure of material damage under irradiation.}
\item\textcolor{black}{ The model for He-induced embrittlement of grain-boundaries described in~\cite{gilbertetal2012} and presented here with revised critical time estimates, while extremely simple and subject to significant assumptions (including grain size and an absence of migration barriers), suggests that He production should not be ignored when designing components for fusion devices.}
\end{itemize}

\textcolor{black}{While accepting the limitations of the materials modeling, the integrated approach in the present study demonstrates the potential to produce engineering relevant predictions -- starting from a reactor design and using a variety of computational tools to arrive at modeling of material properties at the atomic-scale.}

\section*{Acknowledgments}

This work was funded by the RCUK Energy Programme under grant EP/I501045 and the European Communities under the contract of association between EURATOM and CCFE. The views and opinions expressed herein do not necessarily reflect those of the European Commission. This work was carried out within the framework of the European Fusion Development Agreement.




\bibliographystyle{mgilbert_plain_JNM}
\bibliography{he_embrittlement}
\end{document}